\documentclass[preprint,12pt,3p,times,sort&compress]{elsarticle}

\usepackage{lineno,hyperref}
\modulolinenumbers[5]
\usepackage{amsmath}

\newcommand{\mean}[1]{\left\langle #1\right\rangle} 

\journal{Physica A}

\begin{document}

\begin{frontmatter}

\title{The noisy voter model under the influence of contrarians}

\author{Nagi Khalil and Ra\'ul Toral}
\address{IFISC (CSIC-UIB), Instituto de F\'isica Interdisciplinar y Sistemas Complejos, Campus Universitat de les Illes Balears, E-07122 Palma de Mallorca, Spain}




\begin{abstract}
The influence of contrarians on the noisy voter model is studied at the mean-field level. The noisy voter model is a variant of the voter model where agents can adopt two opinions, optimistic or pessimistic, and can change them by means of an imitation (herding) and an intrinsic (noise) mechanisms. An ensemble of noisy voters undergoes a finite-size phase transition, upon increasing the relative importance of the noise to the herding, form a bimodal phase where most of the agents shear the same opinion to a unimodal phase where almost the same fraction of agent are in opposite states. By the inclusion of contrarians we allow for some voters to adopt the opposite opinion of other agents (anti-herding). We first consider the case of only contrarians and show that the only possible steady state is the unimodal one. More generally, when voters and contrarians are present, we show that the bimodal-unimodal transition of the noisy voter model prevails only if the number of contrarians in the system is smaller than four, and their characteristic rates are small enough. For the number of contrarians bigger or equal to four, the voters and the contrarians can be seen only in the unimodal phase. Moreover, if the number of voters and contrarians, as well as the noise and herding rates, are of the same order, then the probability functions of the steady state are very well approximated by the Gaussian distribution. 
\end{abstract}


\end{frontmatter}


\section{Introduction}


Nowadays it is quite common to model the dynamics of opinion as a complex system in terms of agent-based models. In those models ``agents'' or ``units'' can hold different opinions that evolve under dynamical rules that include stochastic effects. In this framework, the global behaviour of the system can be directly linked to the microscopic mechanisms acting at the level of one or a few agents. The voter model (VM) \cite{clsu73,holi75,cafolo09,dozhkodili18} and the majority rule model (MR) by Galam \cite{ga86,ga00,ga02,ga08} are paradigmatic examples of agent-based models where each agent can be in one of two possible opinion states and the dynamics is driven by an imitation process. For the VM a randomly chosen agent blindly copies the state of a neighbour, again randomly chosen, while in the MR a complete group of randomly chosen agents adopt the opinion of the local majority. For finite systems, both models describe an evolution towards a consensus state where all agents share the same opinion \cite{krre03,cafolo09}.

In the real world, however, perfect consensus is an exception and coexistence of opinions is a more likely stable scenario. Both the VM and the MR have been modified in different ways in order to account for this more realistic situation. Among many possibilities, it has been shown that the inclusion of inflexible agents (also known as zealots) or that of contrarians prevents the system from reaching a perfect consensus state, allowing coexistence to prevail. Zealots are agents that never change their opinion, their influence depending both on their number and the detailed structure of the network of interactions\cite{mo03,mopere07,kasaamda14,fues14,moscag15,khsato18,ga07,ga11,crca15}. Contrarians are agents that, contrarily to the imitation rules described above, tend to copy the opposite opinion of a neighbor or to adopt the opinion held by the minority of the group. Their influence on the MR was first studied by Galam \cite{ga04}. He showed that the system can reach two steady states, depending on the concentration of contrarians: if it is small enough, an ordered phase is reached with a majority (but not all) of agents holding one of the two opinions; while for the concentration above a critical value, a disordered phase is reached with the same fraction of agents with different opinions (no majority). See Refs. \cite{coecmamaso02,zhzhzhhu05,maku10,host11,host11a,librhast11,bare13,tama13,ma13,yibazhki13,crblan14,guchlu14,bare15,gacr17,golagoig17,wio}, amongst others, where contrarians have also been considered in other contexts.


Contrarian behaviour has been also studied within the VM, in the mean-field approximation with all agents being neighbours, in Refs. \cite{ba14,ba16}. In the two-role model (TRM) of \cite{ba14,ba16} the agents can choose, at each decission step, between behaving as a ``voter'' and then copy the opinion of a neighbour, or behaving as a ``contrarian'' and adopt the opposite opinion of a neighbour, with given probabilities $1-p$ and $p$, respectively. Observe that the situation is different from the original model by Galam \cite{ga04} where agents have fixed roles and the label of ``voter'' or ``contrarian'' of a single agent remains unchanged during the dynamical evolution. For a system of $N$ agents, three phases can be observed: the bimodal phase if $p<1/(N+1)$, the plain phase for $p=1/(N+1)$, and the unimodal phase for $p>1/(N+1)$. In the bimodal phase, the system keeps most of the time close to the consensus states, where the number of agents holding one particular opinion is much larger than the number holding the opposite one, but the dominant opinion can change with time. Hence, the probability distribution $P(n)$ of the number of agents $n$ holding a particular opinion has maxima at $n=0$ and $n=N$ in the steady state. In the unimodal phase the number of agents in each of the two possible states fluctuates around equal numbers and the distribt¡ution $P(n)$ presents a single maximum at $n=N/2$. In the plain phase, $P(n)$ is uniform for all $n\in[0,N]$.

The situation of the TRM resembles that of the Kirman or noisy voter model \cite{kr93,grma95,catosa15,catosa16}. In that model there are two mechanisms that make an agent to change her binary state: the herding or copying mechanism, as in the VM, and the intrinsic noise allowing agents to change states regardless the sate of the remaining agents. By increasing the relative importance of the noisy with respect to the herding, the system undergoes a finite-size phase transition from a unimodal to a unimodal phase, similar to the one observed in the TRM with respect to the probability $p$.

In this work we study the effect of contrarian agents on the noisy voter model. As in Galam's model, agents are modeled with fixed roles that they keep at all times. The main objective is to unveil the effects of the different mechanisms on the global behaviour of the system, specially on the different phases the system may exhibit, as well as to clarify the similarities and differences between the noisy voter model and the two-role model.


The paper is organized as follows. In the next section we introduce the model and explore some limits, particularly the case of the TRM. Section \ref{sec:3} contains the main results of the system: exact and approximate theoretical expressions are compared against numerical results for simple as well as general cases. Finally, section \ref{sec:4} is devoted to the conclusions. 

\section{Model\label{sec:2}}

The system is made of $N=N_v+N_c$ agents. Each agent can hold any of two possible opinion states. We do not have any particular interpretation in mind, and we will denote the two states generically as ``up'' an ``down''. They could account for ``optimistic'' and ``pessimistic'' states; the ``buy'' and ``sell'' states of brokers in the stock market, or whatever other interpretation. The suffixes $v$ and $c$ stands for the voters and the contrarians, respectively. We assume all agents inside each subgroup to be equivalent, hence the state of the system is fully specify by the set $\{n_v,n_c\}$ of the numbers of up voters $n_v$ and up contrarians $n_c$, with $n\equiv n_v+n_c$ being the total number of up agents. Both voters and contrarians can change their state randomly with certain rates, increasing and decreasing the number $n_v$ and $n_c$. The dynamics implements a Markov chain with the following rates $\pi^{\pm}_{v,c}$ for the allowed transitions:
\begin{itemize}
\item $n_v\to n_v+1$:
 \begin{equation}
 \label{eq:1}
 \pi^+_v(n_v,n_c)=\left(a_v+h_v\frac{n}{N}\right)(N_v-n_v),
 \end{equation}
\item $n_v\to n_v-1$:
 \begin{equation}
 \label{eq:2}
 \pi^-_v(n_v,n_c)=\left(a_v+h_v\frac{N-n}{N}\right)n_v,
 \end{equation}
\item $n_c\to n_c+1$:
 \begin{equation}
 \label{eq:3}
 \pi^+_c(n_v,n_c)=\left(a_c+h_c\frac{N-n}{N}\right)(N_c-n_c),
 \end{equation}
\item $n_c\to n_c-1$:
 \begin{equation}
 \label{eq:4}
 \pi^-_c(n_v,n_c)=\left(a_c+h_c\frac{n}{N}\right)n_c.
 \end{equation}
\end{itemize}
The rates depend on the ``noise'' coefficients $a_v$ and $a_c$ and the ``herding'' $h_v$ and ``anti-herding'' $h_c$ coefficients. According to the chosen rates, a voter and a contrarian can change their states regardless the state of the other agents with rates $a_v$ and $a_c$, respectively. Moreover, a voter can also change her state with a rate proportional to the total number of agents holding the opposite state than hers, while a contrarian changes state with a rate proportional to the number of agents holding her state. In terms of pairwise interactions, the herding mechanism in equivalent to an agent copying the state of another agent selected randomly with rate $h_v$ from the whole population, while the anti-herding mechanism is equivalent to an agent copying the opposite state of another agent again selected randomly with rate $h_v$ from the whole population. Observe that the rates at Eqs. (\ref{eq:1})--(\ref{eq:4}) are quadratic functions of $n_v$ and $n_c$. However, $\pi^+_{v,c}+\pi^-_{v,c}$ are linear functions, a property that makes the system analytically solvable, see the discussion of Ref.~\cite{scsp88} addressing the case of the VM. 

It is useful to rewrite the contrarian rates as that of the voter, as follows:
\begin{equation}
 \label{eq:5}
 \pi^+_c(n_v,n_c)=\left(\bar a_c+\bar h_c\frac{n}{N}\right)(N_c-n_c),
\end{equation}
\begin{equation}
 \label{eq:6}
 \pi^-_c(n_v,n_c)=\left(\bar a_c+\bar h_c\frac{N-n}{N}\right)n_c,
\end{equation}
with
\begin{equation}
 \label{eq:7}
 \bar a_c\equiv a_c+h_c
\end{equation}
and 
\begin{equation}
 \label{eq:8}
 \bar h_c\equiv -h_c.
\end{equation}
That is, the contrarians can be seen as noisy voters with negative herding parameter 
\begin{equation}
 \label{eq:9}
 \bar h_c\le 0, 
\end{equation}
but with a noisy coefficient greater of equal to $-\bar h_c$
\begin{equation}
 \label{eq:110}
 \bar a_c \ge -\bar h_c
\end{equation}
which insures the total rates to be non negative. As, similarly, a contrarian with negative herding acts as a voter, and in order to avoid inconsistent notation (contrarians behaving as voters or viceversa), we consider, from now on, that all herding parameters are non negative $h_c,h_v\ge 0$.  By construction, the noise coefficients can not be negative either, $a_v,a_c\ge0$.

Agents can also adopt the radical form of ``zealots''. They are agents that never change their state. Formally a zealot is either a voter or a contrarian with $a_{c,v}=h_{v,c}=0$, so the rate of changin state is always zero. A zealot's opinion is determined solely by its initial state.

\subsection{The case of the two-role model}

Suppose a system of $N$ equivalent agents that can adopt the role a voter with probability $1-p$ and the role of a contrarian with probability $p$, as the TRM considered in Ref.~\cite{ba14}. The possibility of an intrinsic change of state is here disregarded. Then the rates for the allowed transitions, using the structure of the rates (\ref{eq:1})--(\ref{eq:4}), read
\begin{equation}
 \label{eq:11}
 \pi^+(n)=h\left[(1-p)\frac nN+p\frac{N-n}{N}\right](N-n),
\end{equation}
\begin{equation}
 \label{eq:12}
 \pi^-(n)=h\left[(1-p)\frac{N-n}{N}+p\frac{n}{N}\right]n,
\end{equation}
where $h$ is a constant. Notice a slight difference of our rates if compared to that of Ref.~\cite{ba14}: in our approach, when an agent adopts the role of a contrarian, her state changes taking into account also her own state. The difference disappears for $N\gg 1$. 

After a rearrangement, the rates (\ref{eq:11}) and (\ref{eq:12}) can be written as
\begin{equation}
 \label{eq:13}
 \pi^+(n)=\left[hp+h(1-2p)\frac nN\right](N-n),
\end{equation}
\begin{equation}
 \label{eq:14}
 \pi^-(n)=\left[hp+h(1-2p)\frac{N-n}{N}\right]n,
\end{equation}
which are a particular case of (\ref{eq:1})--(\ref{eq:2}) if $p<1/2$, and of (\ref{eq:5})--(\ref{eq:6}) for $p>1/2$.

For all said, the TRM studied in \cite{ba14,ba16} is a particular case (for $N\gg 1$) of the general model considered here. On the one hand, if the probability $p$, for a an agent of the TRM to be a contrarian, is smaller than $1/2$ then the model is equivalent to an ensemble of noisy voters with noise $hp$ and herding $h(1-2p)$. On the other hand, if $p>1/2$ then the equivalence is with an ensemble of noisy contrarians with noise $hp$ and herding $h(2p-1)$. Finally, for $p=1/2$ we have a random walk with transition rate $h/2$.

Once we have established the relationship between the two-role model of Ref.~\cite{ba14} and our general model of voters and contrarians, we analyze the latter in detail and determine the possible phases and transitions that it can present. Our intention is to identify whether consensus or coexistence of opinions as a function of the system parameters and relative numbers of voters and contrarians. We will also study the role of zealots in the final outcome of the dynamics.

\section{Theory\label{sec:3}}

At the mesoscopic level, the fundamental quantity is the probability $P(n_v,n_c,t)$ of finding the system in state $\{n_v,n_c\}$ at time $t$. It satisfies the following master equation~\cite{vanKampen:2007,Toral-Colet:2014}
\begin{equation}
 \label{eq:15}
 \frac{\partial}{\partial t}P(n_v,n_c,t)=\sum_{k\in\{v,c\}}\sum_{s\in\{+,-\}}(E_k^s-1)\pi^{-s}_k(n_v,n_c)P(n_v,n_c,t),
\end{equation}
where $E_v^\pm$ and $E_c^\pm$ are the step operators defined such that $E_v^\pm f(n_v,n_c)=f(n_v\pm 1,n_c)$ and $E_c^\pm f(n_v,n_c)=f(n_v,n_c\pm 1)$ for any function $f(n_v,n_c)$. Equations for the moments 
\begin{equation}
 \label{eq:16}
 M_{ij}\equiv \mean{n_v^in_c^j}=\sum_{n_v}\sum_{n_c}n_v^in_c^jP(n_v,n_c,t),\qquad i,j\in\mathbb N,
\end{equation}
where $\mean{\cdot}$ denotes an average over the probability function, can be easily inferred form Eq. (\ref{eq:15}) by multiplying it by $n_v^in_c^j$ and summing over all possible values of $n_v$ and $n_c$. After some algebra, we arrive at 
\begin{equation}
 \label{eq:17}
 \frac{dM_{ij}}{dt}=\sum_{k\in\{v,c\}}\sum_{s\in\{+,-\}}\mean{\pi_k^s(E_k^s-1)n_v^in_c^j}.
\end{equation}
 To obtain this equation, the following property has been used 
\begin{equation}
 \label{eq:18}
 \mean{f(n_v,n_c)(E_k^s-1)g(n_v,n_c)}=\mean{g(n_v,n_c)(E_k^{-s}-1)f(n_v,n_c)},
\end{equation}
valid for any pair of functions $f$ and $g$

It turns out that the hierarchy of equations (\ref{eq:17}) can be closed at any order (value of $i+j$) because $\sum_{s\in\{+,-\}}\mean{\pi_k^s(E_k^s-1)n_v^in_c^j}$ involves only moments of degree $i+j$ or less, a direct consequence of the form of the rates. Namely, $(E_k^s-1)n_v^in_c^j$ is either zero or a polynomial of degree $i+j-1$ whose leading coefficient has different signs for different values of $s$. Hence the sum on the r.h.s. of Eq. (\ref{eq:17}) is either zero or it involves the combination of $\pi_k^+-\pi_k^-$ times a polynomial of degree $i+j-1$. Since $\pi_k^+-\pi_k^-$ is of degree one for the rates (\ref{eq:1})--(\ref{eq:4}), we obtain the desired result. 

We consider next the steady--state solutions of Eqs. (\ref{eq:17}). But first, it is more natural to take the partial and global magnetizations, defined as 
\begin{equation}
 \label{eq:19}
 x_v=\frac{2n_v}{N_v}-1; \qquad x_c=\frac{2n_c}{N_c}-1; \qquad x=\frac{2n}{N}-1=x_c+x_v.
\end{equation}
The new quantities take values in $[-1,1]$, and are correlated in general. It is not difficult to see that the steady-state values of the moments of degree one are
\begin{equation}
 \label{eq:19b}
 \mean{n_v}=\frac{N_v}{2} \Leftrightarrow \mean{x_v}=0, 
\end{equation}
\begin{equation}
 \label{eq:20}
 \mean{n_c}=\frac{N_c}{2} \Leftrightarrow \mean{x_c}=0, 
\end{equation}
in agreement with the symmetry of the problem. In a similar way, we can obtain the second moments. However, the explicit expression are very long, and are given in \ref{ap1}.

The knowledge of the second moments can be used to infer the phase of the system, through the following result: 

\noindent\textbf{Lemma}. \emph{Let the probability function $P(x): D\rightarrow \mathbb R^+$ be an even function, monotonic in $D^+$, with $D=\{x_i\equiv -1+2i/N, \ i=0,\dots,N\}$ and $D^+=\{x\in D| x> 0\}$. Then $P(x)$ is non-decreasing (resp. non-increasing) in $D^+$ if and only if $\mean{x^2}\ge (\text{resp.} \le) \frac{N+2}{3N}$. The equality holds when $P(x)$ is constant.} 

\noindent{\textbf{Proof of the lemma}. Under the hypothesis, $P(x)$ can be either (a) non-decreasing and non-constant, (b) non-increasing and non-constant, or (c) constant. Case (c) is evident. In case (a) it is $P(x_i)\le P(x_{i+1})$ for $i=0,\dots, N-1$. Take $\tilde P(x_i)=P(x_i)-\frac{1}{N+1}$, then $\tilde P(x_i)\le \tilde P(x_{i+1})$, and there are two numbers $i_m<i_M$ so that $P(x_i)=0$ for $i_m+1\le i\le i_M-1$. Moreover, $P(x_i)<0\ (>0)$ for $i\le i_m\ (i\ge i_M)$, and $k\equiv -\sum_{i\le i_m}\tilde P(x_i)=\sum_{i\ge i_M}\tilde P(x_i)>0$. Hence, $\mean{x^2}-\frac{N+2}{3N}=\sum_{i=0}^N x_i^2\tilde P(x_i)=\sum_{i\le i_m}x_i^2\tilde P(x_i)+\sum_{i\ge i_M}x_i^2\tilde P(x_i)\ge x_{i_m}^2\sum_{i\le i_m}\tilde P(x_i)+x_{i_M}^2\sum_{i\ge i_M}\tilde P(x_i)=k(x_{i_M}^2-x_{i_m}^2)>0$. Similarly, we can prove (b). \textbf{End of proof.}

By symmetry considerations, $P(x_v)$, $P(x_c)$, and $P(x)$ are even functions. In addition, if we can prove that they are monotonic functions when their arguments are positive, then it follows that there are three possible phases for the subsystem of voters and contrarians: the bimodal phase (with $x=0$ being the less probable value), the unimodal phase (with $x=0$ being the most probable value), and the plain phase as the border case. Next, we consider some cases separately.

\subsection{The noisy voter model}

When $N_c=0$, we recover the noisy voter model. For this case, it was proven \cite{khsato18} that the probability function $P(x)$ for $x>0$ is monotonic, and we can use the previous lemma. The second moment is 
\begin{equation}
 \label{eq:22}
 \mean{x_v^2}=\frac{2a_v+h_v}{2Na_v+h_v},
\end{equation}
so for $a_v/h_v< 1/N$, i.e. $\mean{x_v^2}> \frac{N+2}{3N}$, the system is in the bimodal phase; for $a_v/h_v>1/N$, i.e. $\mean{x_v^2}< \frac{N+2}{3N}$, the system is in the unimodal phase; and for $a_v/h_v=1/N$, i.e. $\mean{x_v^2}= \frac{N+2}{3N}$, the system is in the plain phase, as shown in Fig. \ref{fig:1}, see \cite{khsato18} for further details. 

\begin{figure}[!h]
 \centering
 \includegraphics[width=.45\linewidth]{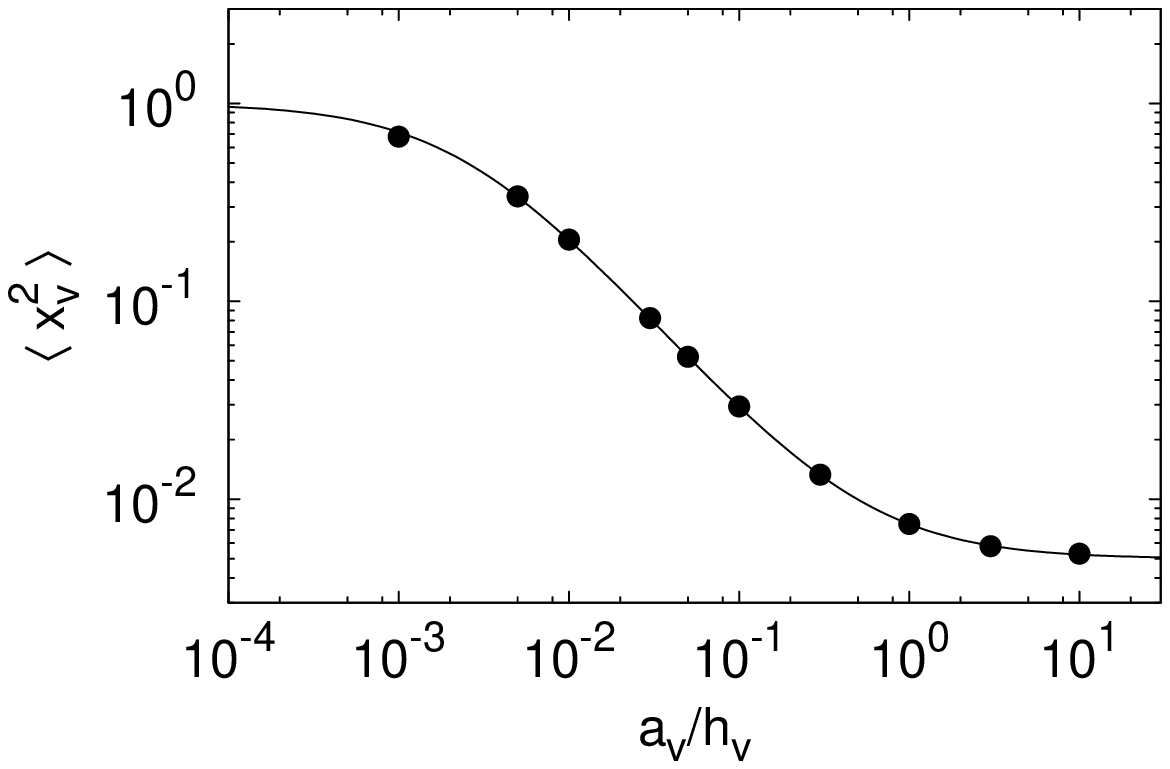}
 \includegraphics[width=.45\linewidth]{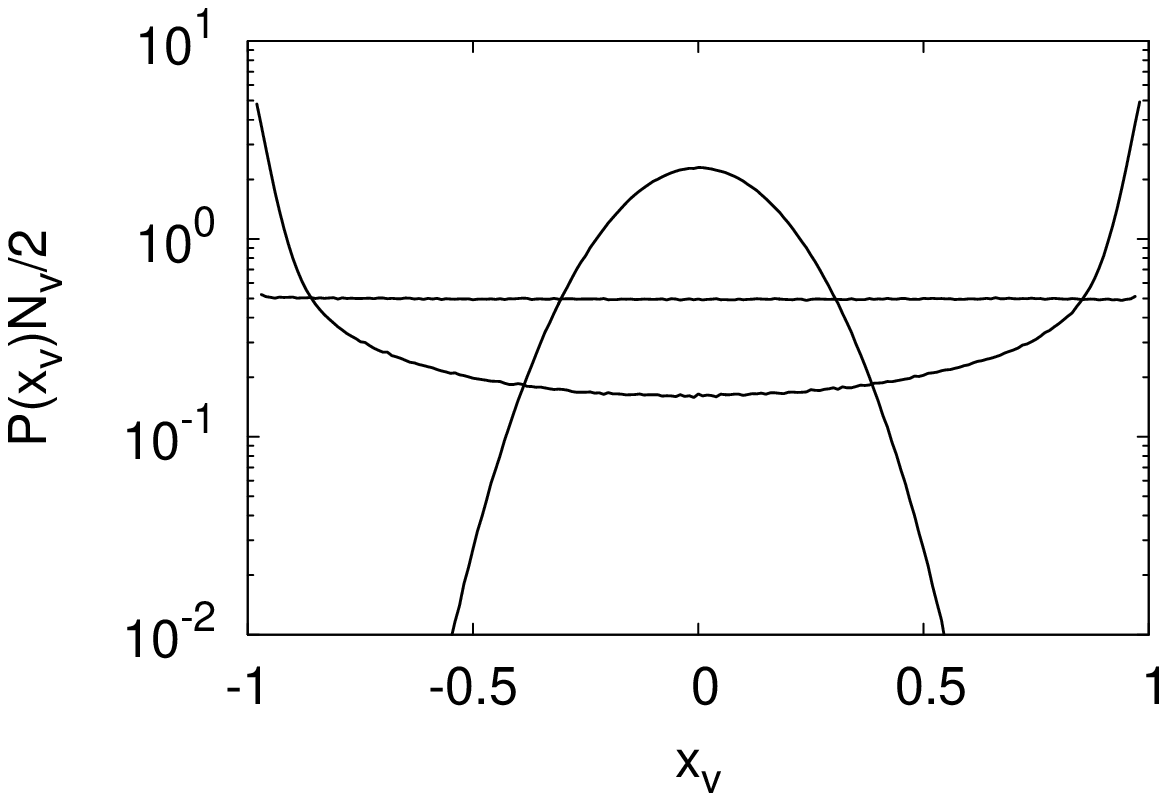}
 \caption{Second moments (left) and probability functions (right) for the magnetization of a system of $N_v=200$ voters and no contrarians $N_c=0$. Left: the line is from Eq. (\ref{eq:22}) and symbols are from the numerical simulations. Right: numerical results for $a_v/h_v=1/(5N_v)$ (convex bimodal function), $1/N_v$ (plain function), and $20/N_v$ (concave unimodal function) showing the the possible three phases.}
 \label{fig:1}
\end{figure}

A simple modification of this case, allows us to consider the influence of zealots in a system of voters. Formally, we identify zealots as a set of $N_c> 0$ contrarian agents with zero rates, $a_c=h_c=0$. The state of the zealot is determined by the initial condition and it remains unchanged during the evolution of the system. The distribution of the states of the voters then depends strongly on the number and states of the zealots. If the number of zealots of opposite states are the same, the so called balanced case, the bimodal and plain phases in the distribution of voters $P(n_c)$ always disappear. If the number of opposite zealots are different (unbalanced case) the system loses its optimistic-pessimistic symmetry and two new phases appear (see \cite{khsato18} for details): an extreme asymmetric (EA) phase (where the maximum of the probability distribution occurs at the consensus value favored by the zealots) and an asymmetric unimodal (AU) phase where the maximum, being still tilted towards the zealot-favored opinion, is located far from the extreme consensus state.

\subsection{Only contrarians}

We now consider the case of only contrarians, i.e. $N_v=0$. The master equation for $P(n)$ with $n=n_c$ can be easily written down, and from it we deduce the following useful relation for the steady-states probability functions, see appendix C of \cite{khsato18}, 
\begin{equation}
 \label{eq:23}
 P(n)=\frac{\pi^+_c(n-1)}{\pi^-_c(n)}P(n-1), \qquad n=1,\dots, N.
\end{equation}
Using the rates (\ref{eq:3}) and (\ref{eq:4}) we have 
\begin{equation}
 \label{eq:24}
 \frac{\pi^+_c(n-1)}{\pi^-_c(n)}=\frac{\left(a_c+h_c\frac{N-n+1}{N}\right)(N-n+1)}{\left(a_c+h_c\frac{n}{N}\right)n}
\end{equation}
which is smaller than $1$ for $n>N/2$ and greater than $1$ for $n<N/2$. Namely, the steady-state probability function verifies the hypothesis of the lemma. Moreover, $n=N/2$ is always a global maximum, and the only possible phase is the unimodal one. The same conclusion can be reached by using the lemma: the exact expression for the second moment is
\begin{equation}
 \label{eq:25}
 \mean{x_c^2}=\frac{2a_c+h_c}{2Na_c+(2N-1)h_c},
\end{equation}
Using $a_c\ge 0,\,N>1$, it follows immediately that $\mean{x_c^2}<(N+2)/(3N)$, meaning (according to the lemma) that the only possible phase is the unimodal one. 

The expression in Eq. (\ref{eq:25}) is an increasing function of the noise coefficient $a_c$, see Fig. \ref{fig:2}, meaning that the anti-herding (contrarian) mechanism is more efficient than the noise to lead the system deep inside the unimodal phase, since the smaller the value of $a_c/h_c$ the smaller the value of $\mean{x_c^2}$. The opposite behaviour is observed for the noisy voter model, i.e. Eq. (\ref{eq:22}) is a decreasing function of $a_v/h_v$. For $a_c\gg h_c$ we get $\mean{x_c^2}\simeq 1/N$, the anti-herding mechanism does not act, and the behaviour of the system is the same as the noisy voter model for $a_v=a_c\gg h_v/N_v$, as expected. 

\begin{figure}[!h]
 \centering
 \includegraphics[width=.45\linewidth]{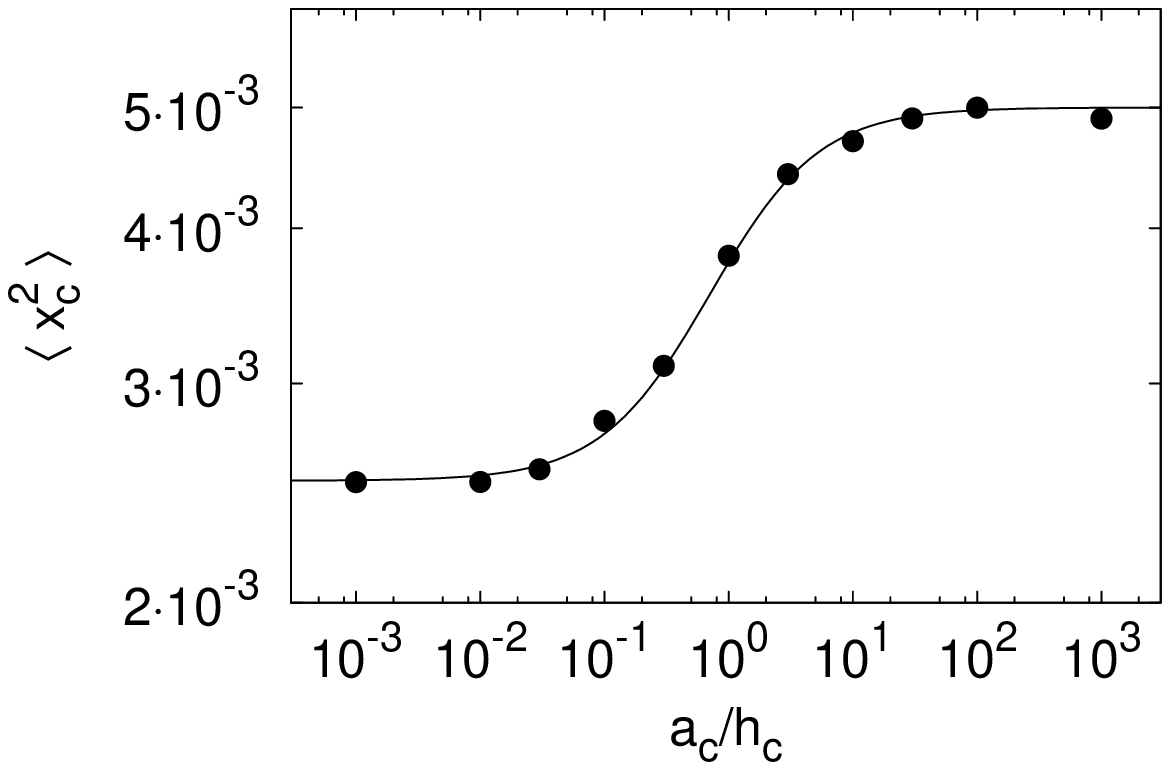}
 \includegraphics[width=.45\linewidth]{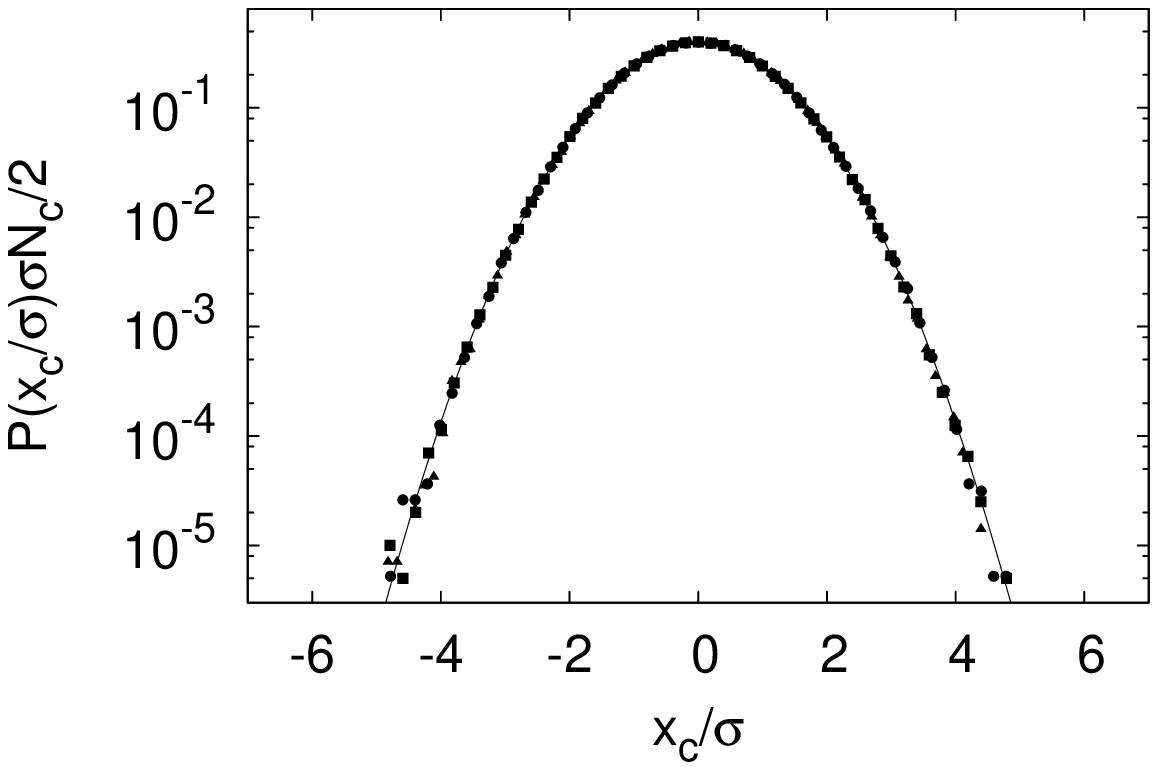}
 \caption{Second moments (left) and probability functions (right) for a system of $N_c=200$ contrarians and no voters $N_v=0$. Left: the line is from Eq. (\ref{eq:25}) and symbols are from the numerical simulations. Right: numerical results for $a_v/h_v=10^{-3}$ (squares), $10^{-1}$ (circles), and $10^2$ (triangles). The data has been scaled with their variance $\sigma\equiv\sqrt{\mean{x_c^2}}$, and the line is a Gaussian distribution with unit variance.}
 \label{fig:2}
\end{figure}

It is also possible to study the effect that zealots have in a system of contrarians. Similarly to the previous subsection we just need to consider a set of $N_v>0$ voters with $a_v=h_v=0$, that will act as zealots. In the balanced case of an even number $N_v$ of zealots, with the same number of zealots having opposite states, i.e. $n_v=N_v/2$, we can consider the effect that zealots have on the contrarian rates, rewriting Eqs.(\ref{eq:3},\ref{eq:4}) as: 
\begin{eqnarray}
 \label{eq:26}
 \pi^+_c(n_v,n_c)&=&\left(a_c+h_c\frac{N-n}{N}\right)(N_c-n_c)=\left(\tilde{a}_c+\tilde{h}_c\frac{N_c-n_c}{N_c}\right)(N_c-n_c),\\
 \label{eq:27}
 \pi^-_c(n_v,n_c)&=&\left(a_c+h_c\frac{n}{N}\right)n_c=\left(\tilde{a}_c+\tilde{h}_c\frac{n_c}{N_c}\right)n_c,
\end{eqnarray}
with effective noise and herding coefficients,
\begin{eqnarray}
\tilde{a}_c&=&a_c+h_c\frac{N_v}{2N,}\\
\tilde{h}_c&=&\frac{N_ch_c}{N},
\end{eqnarray}
meaning that the roles of zealots on a set of contrarian agents is to increase the effective noise and to decrease the anti-herding constant. As a consequence, the system keeps always in the unimodal phase. 

\subsection{Noisy voters under the influence of extreme contrarians}
We consider in this subsection a more general case with a mixture of voters and contrarians. Moreover, since we have already seen that the noise mechanism diminishes the contrarian effect, from now on we set $a_c=0$, so that the effect of the contrarians will be tuned only through changing $h_c$ and $N_c$. This way, the relevant free parameters become $a_v/h_v$, $h_c/h_v$, $N_v$, and $N_c$. A contrarian with $a_c=0$ never changes spontaneously its state and will be termed as a ``extreme`` contrarian.

Take the subsystem of voters. From the exact expression of $\mean{x_v^2}$, see \ref{ap1}, we can compute the values of the parameters where the different phases appear, as previously discussed. By imposing $\mean{x_v^2}=\dfrac{N_v+2}{3N_v}$ we obtain the critical values of $a_v=a_v(h_c/h_v,N_v,N_c)$ as the (positive) solution of the quadratic equation 
\begin{equation}
 \label{eq:28}
 A a_v^2+B a_v+C=0,
\end{equation}
where 
\begin{eqnarray}
 \label{eq:29}
A&=&2(N_v+N_c) (2N_c+N_v-1)>0,\\
 \label{eq:30}
 B&=&N_v^2h_c/h_v+2(2N_c-1)[(h_c/h_v+1)N_c-1]+[(4h_c/h_v+3)N_c-3-h_c/h_v]N_v,\\
 \label{eq:31}
 C&=&1+h_c/h_v(N_c-1) (N_v+2N_c-1)+(N_c-4)N_c.
\end{eqnarray}
As $A,B>0$, a positive solution to Eq. (\ref{eq:28}) requires $C<0$. If $C>0$ all solutions will be negative and, effectively, there are no transitions between different phases, being the unimodal the only possible phase. The limiting case happens when $C=0$, or
\begin{equation}
 \label{eq:32}
 \frac{h_c}{h_v}=\frac{(4-N_c)N_c-1}{(N_c-1)(N_v+2N_c-1)}.
\end{equation}
As $h_c,h_v\ge 0$, this condition can only be fulfilled if $(4-N_c)N_c\ge1$, or $N_c=1,2,3$. In summary, for $N_c\ge4$ the only possible outcome is the unimodal distribution of up voters. On the other hand, for $N_c=0,1,2,3$ it is possible to switch from the unimodal to the bimodal phase, passing through the plain phase, by varying the parameters $a_v,h_v,h_c$. The phase diagram is sketched in  Fig. \ref{fig:3}.

\begin{figure}[!h]
 \centering
 \includegraphics[width=.32\linewidth]{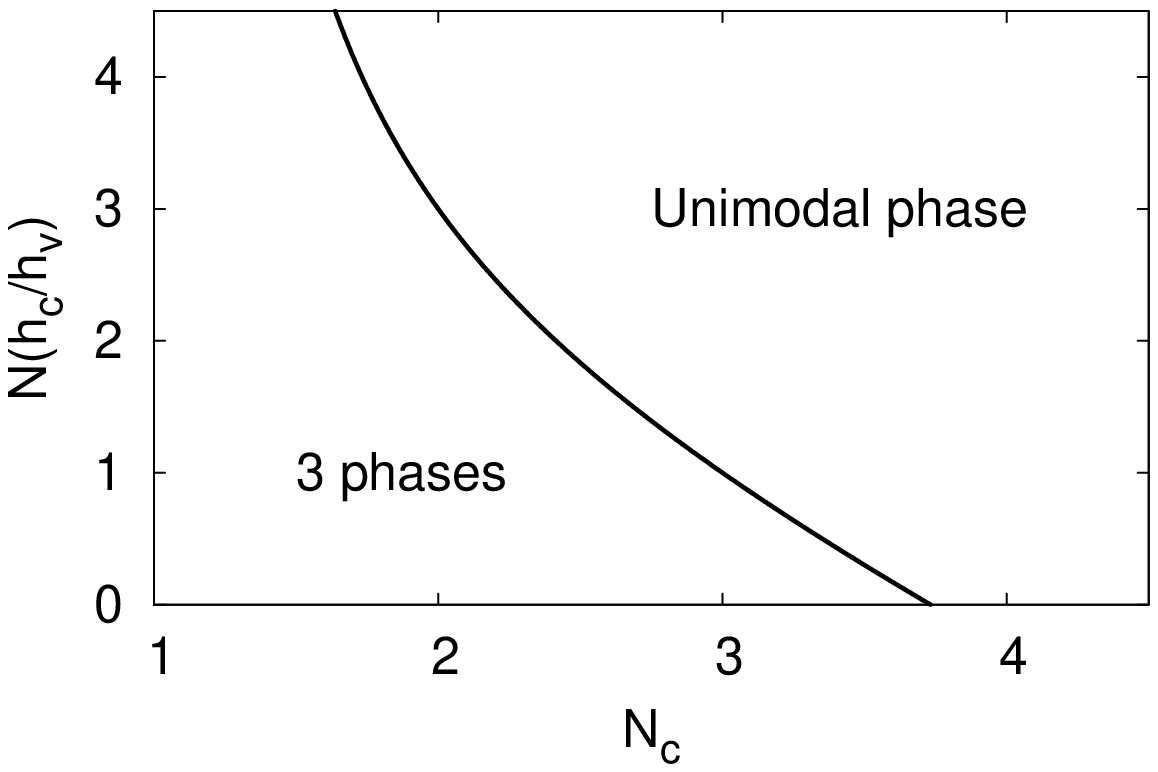}
 \includegraphics[width=.32\linewidth]{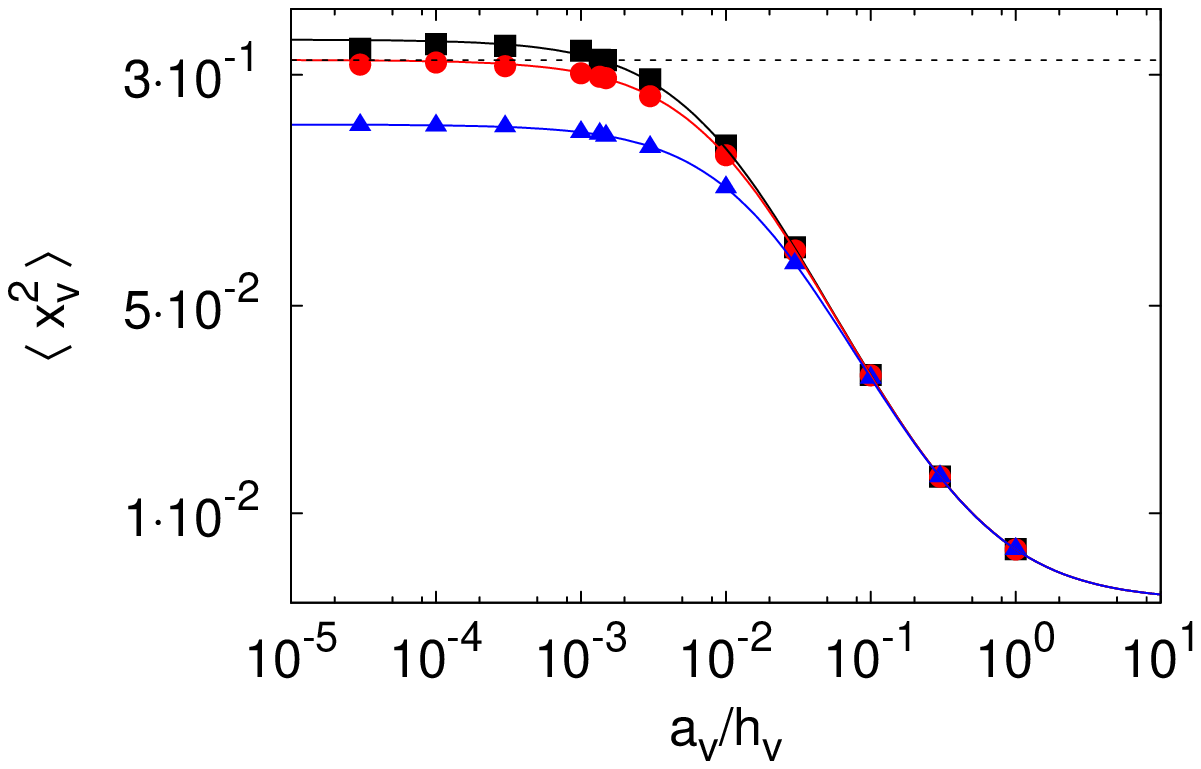}
 \includegraphics[width=.32\linewidth]{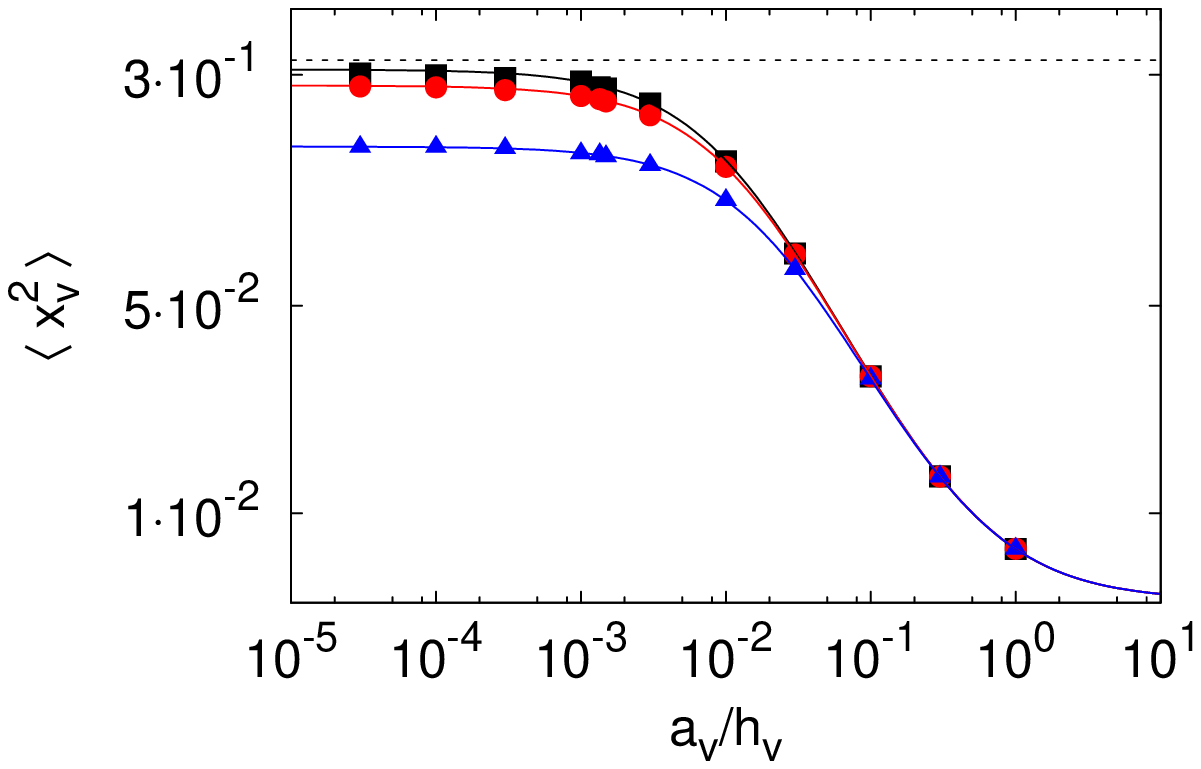}
 \caption{Left: A phase diagram with two regions: below the critical line the voters can be in the three phases, depending on the value of $a_v$, while above the critical line only the unimodal phase is possible. The exact position of the critical line depends on the value of $N_v$, however this dependence is through the prefactor $N/(N+N_c-1)\simeq 1$ for $N\gg 1$. Center and right: Numerical results (symbols) and theory (lines) for the second moment of the magnetization of a system of voters under the influence of extreme contrarians as a function of the noisy to herding ratio $a_v/h_v$, and for $h_c/h_v/(h_c/h_v)_0=10^{-1}$ (black square), $1$ (red circles), and $10$ (blue triangle), with $(h_c/h_v)_c$ given by Eq. (\ref{eq:32}). Center: $N_v=197$ and $N_c=3$. Right: $N_v=196$ and $N_c=4$. The dashed horizontal line is the value $\mean{x_v^2}=(N_v+2)/(3N_v)$ for which the bimodal-unimodal transition occurs.}
 \label{fig:3}
\end{figure}

Let us consider that $N_c\le 3$, so the voters can be in the bimodal phase. If we now take the limit of $h_c/h_v\to 0$, one could naively expect that the contrarians behave like zealots. But this is not the case, as can be seen from Eq. (\ref{eq:32}). Take, for instance, two contrarians $N_c=2$ that initially have opposite opinions. For $h_c/h_v=0$, i.e. the contrarians being zealots, voters are always in the unimodal phase, as already said in a previous subsection. However, for $0<h_c/h_v<3/(N_v+3)$ the voters are in the bimodal phase, even if $h_c/h_v\ll 1$. The singularity of the limit $h_c/h_v\to 0$ has to do with the time needed for the system to reach the steady state. While for $h_c/h_v=0$ the relaxing time is of order of $(a_v+h_v)^{-1}$, for $h_c/h_v\ll a_v/h_v$ it is or order of $h_c^{-1}$.

Another important observation concerns the region of existence of the bimodal phase that disappears for $N_v,N_c\to \infty$ (finite-size character of the transitions). For the case of the noisy voter model, both with and without the influence of contrarians (and zealots), the bimodal phase disappears with a small amount of noise $a_v\sim h_v/N$, a small number of contrarians $N_c/N_v\sim 4/N_v$, and/or a small value of the anti-herding parameter $h_c\sim h_v/N_v$. For $N_c\sim N_v$ both subgroups of agents are typically very deep in the unimodal phase. 

\subsection{The unimodal phases }
We have already seen that the most common phase for both the voters and the contrarians is the unimodal phase. This is very apparent if the constants are the same for both kind of agents, and/or $N_c\sim N_v\gg 1$. Moreover, if both kind of agents are deep inside the unimodal phase, e.g. if the noise is big enough, then we expect the two communities to decouple, in the sense specified below. 

First, we derive an equation for the marginal probability functions $P(n_v)$ and $P(n_c)$ from the master equation (\ref{eq:15}). By summing Eq. (\ref{eq:15}) over all possible values of $n_c$, we get the following equation
\begin{equation}
 \label{eq:33}
 \frac{\partial}{\partial t}P(n_v,t)=\sum_{s\in\{+,-\}}(E_v^s-1)\sum_{n_c=0}^{N_c}\pi^{-s}_v(n_v,n_c)P(n_v,n_c,t)
\end{equation}
and a similar one for $P(n_c,t)$. Now we introduce the fundamental assumptions 
\begin{equation}
 \label{eq:34}
 \sum_{n_c}\pi^{-s}_v(n_v,n_c)P(n_v,n_c,t)\simeq \pi^{-s}_v(n_v,\mean{n_c})P(n_v,t)
\end{equation}
and the analogous one for the contrarians. This way, we obtain the following system of equations:
\begin{equation}
 \label{eq:35}
 \frac{\partial}{\partial t}P(n_v,t)\simeq \sum_{l\in\{+,-\}}(E_v^l-1)\pi^{-l}_v(n_v,\mean{n_c})P(n_v,t),
\end{equation}
\begin{equation}
 \label{eq:36}
 \frac{\partial}{\partial t}P(n_c,t)\simeq \sum_{l\in\{+,-\}}(E_c^l-1)\pi^{-l}_c(\mean{n_v},n_c)P(n_c,t).
\end{equation}
For the steady-state solutions, and due to the up-down symmetry, the two equations decouple one from the other. More specifically, since $\mean{n_v}=N_v/2$ and $\mean{n_c}=N_c/2$, the rates become
\begin{equation}
 \label{eq:37}
 \pi^{+}_v(n_v,\mean{n_c})=\left(\tilde a_v+\tilde h_v\frac{n_v}{N_v}\right)(N_v-n_v); \quad 
 \pi^{-}_v(n_v,\mean{n_c})=\left(\tilde a_v+\tilde h_v\frac{N_v-n_v}{N_v}\right)n_v,
\end{equation}
where $\tilde a_v=a_v+\frac{N_c}{2N}h_v$ and $\tilde h_v=\frac{N_v}{N}h_v$, and 
\begin{equation}
 \label{eq:38}
 \pi^{+}_c(\mean{n_v},n_c)=\left(\tilde a_c+\tilde h_c\frac{N_c-n_c}{N_c}\right)(N_c-n_c); \quad 
 \pi^{-}_c(\mean{n_v},n_c)=\left(\tilde a_c+\tilde h_c\frac{n_c}{N_c}\right)n_c,
\end{equation}
with $\tilde a_c=a_c+\frac{N_v}{2N}h_c$ and $\tilde h_c=\frac{N_c}{N}h_c$. The respective second moments become, after using Eqs. (\ref{eq:22}) and (\ref{eq:25}), 
\begin{equation}
 \label{eq:39}
 \mean{x_v^2}\simeq \frac{2\tilde a_v+\tilde h_v}{2N_v\tilde a_v+\tilde h_v}=\frac{2a_v+h_v}{2N_va_v+\frac{N_v(N_c+1)}{N}h_v}\underset{a_v\ge 0}{\le} \frac{N_v+N_c}{N_v(1+N_c)}\underset{N_c\ge 2}{\le} \frac{N_v+2}{3N_v},
\end{equation}
and 
\begin{equation}
 \label{eq:40}
 \mean{x_c^2}\simeq\frac{2\tilde a_c+\tilde h_c}{2N_c\tilde a_c+(2N_c-1)\tilde h_c}=\frac{1}{N_c}\frac{2a_c+h_c}{2a_c+\frac{N+N_c-1}{N}h_c}\le \frac{1}{N_c}\le \frac{N_c+2}{3N_c}.
\end{equation}
See Fig. \ref{fig:4} where we compare the latter approximate expression with exact ones and numerical results. 

\begin{figure}[!h]
 \centering
 \includegraphics[width=.32\linewidth]{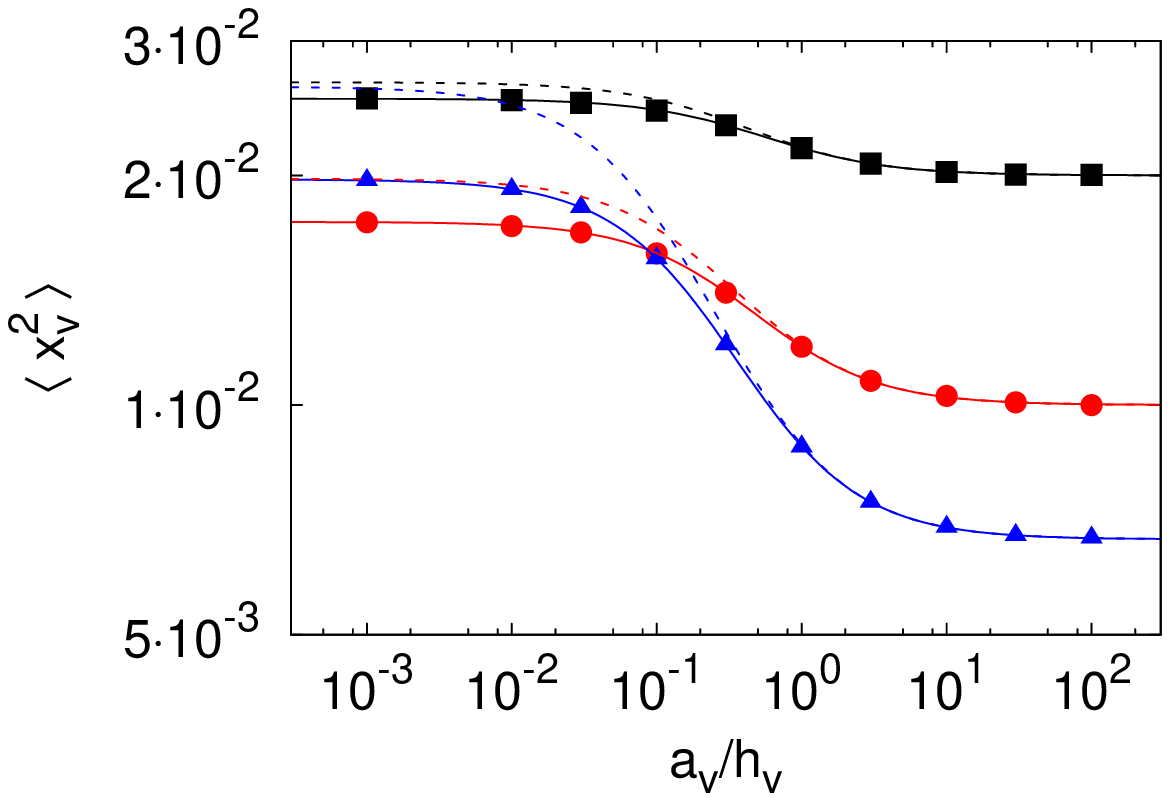}
 \includegraphics[width=.32\linewidth]{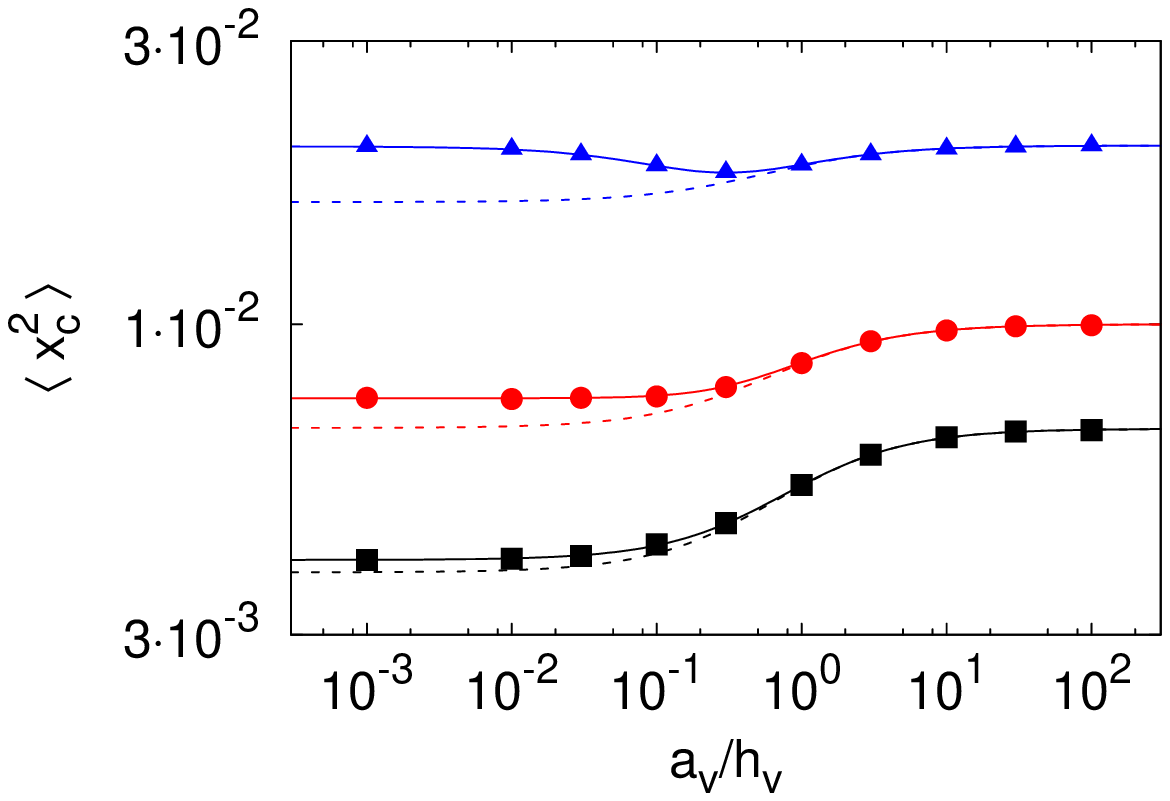}
 \includegraphics[width=.32\linewidth]{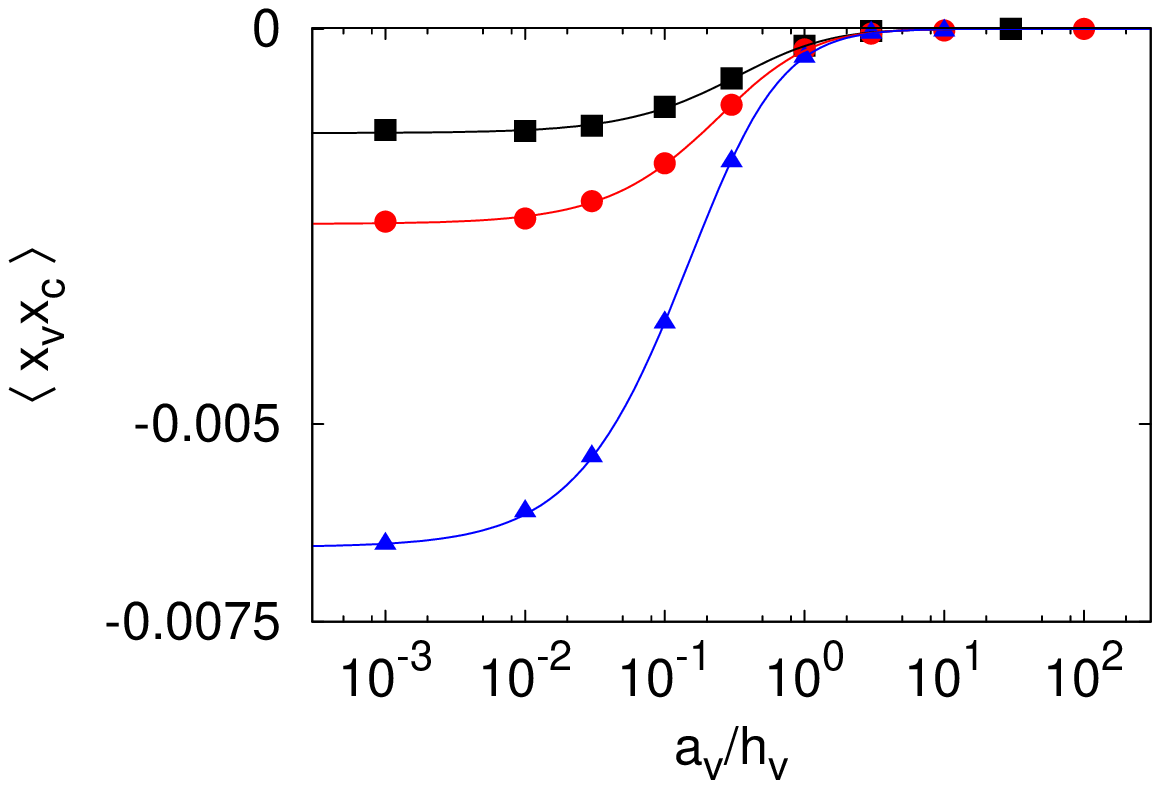}
 \caption{Numerical (symbols), exact (solid lines), and approximate results of Eqs. \eqref{eq:39} and \eqref{eq:40} (dashed lines) for the second moments of the magnetization of a system of $200$ agents, and $N_v=50$ (black squares), $100$ (red circles), and $150$ (blue triangle) voters as a function of noisy to herding ratio. Voters and contrarians have the same rates, $a_v=a_v$ and $h_v=h_c$. }
 \label{fig:4}
\end{figure}

We can also obtain approximate expressions for the probability functions. Following the well-known approach of \cite{vanKampen:2007}, if the numbers of voters $N_v$ and contrarians $N_c$ are large, then the distributions are solution to the Fokker-Planck equations resulting form Taylor expanding the r.h.s of the master equations (\ref{eq:35}) and (\ref{eq:36}) up to second order in $1/N_v$ and $1/N_c$. For the steady-state solutions, the equations are
\begin{equation}
 \label{eq:41}
 \frac{d}{dx_v}\left[2\tilde a_v x_vP(x_v)\right]+\frac{1}{2N}\frac{d^2}{dx_v^2}\left\{\left[4\tilde a_v+2\tilde h_v(1-x_v^2)\right]P(x_v)\right\}=0,
\end{equation}
which was obtained previously in \cite{alluwa08}, and
\begin{equation}
 \label{eq:42}
 \frac{d}{dx_c}\left[2(\tilde a_c+\tilde h_c) x_cP(x_c)\right]+\frac{1}{2N}\frac{d^2}{dx_c^2}\left\{\left[4\tilde a_c+2\tilde h_c(1+x_c^2)\right]P(x_c)\right\}=0.
\end{equation}
The corresponding solutions are 
\begin{equation}
 \label{eq:43}
 P(x_v)=\frac{1}{Z_v}\left(1-\frac{1}{2N_v\tilde a_v/\tilde h_v+1}\frac{x_v^2}{\mean{x_v^2}}\right)^{\frac{N_v\tilde a_v}{\tilde h_v}-1}\underset{N_v\tilde a_v/\tilde h_v\gg 1}{\longrightarrow}\frac{\exp\left(-\frac{x_v^2}{2\mean{x_v^2}}\right)}{\sqrt{2\pi \mean{x_v^2}}},
\end{equation}
and
\begin{equation}
 \label{eq:43b}
 P(x_c)=\frac{1}{Z_c}\left[1+\frac{1}{2\left(\tilde a_c/\tilde h_c+1\right)N_c-1}\frac{x_c^2}{\mean{x_c^2}}\right]^{-\left(\tilde a_c/\tilde h_c+1\right)N_c-1}\underset{N_c(\tilde a_c/\tilde h_c+1)\gg 1}{\longrightarrow}\frac{\exp\left(-\frac{x_c^2}{2\mean{x_c^2}}\right)}{\sqrt{2\pi \mean{x_c^2}}},
\end{equation}
where $Z_v$ and $Z_c$ are normalization constants. The Gaussian approximations are verified in Fig. \ref{fig:6} for one representative case. 

\begin{figure}[!h]
 \centering
 \includegraphics[width=.45\linewidth]{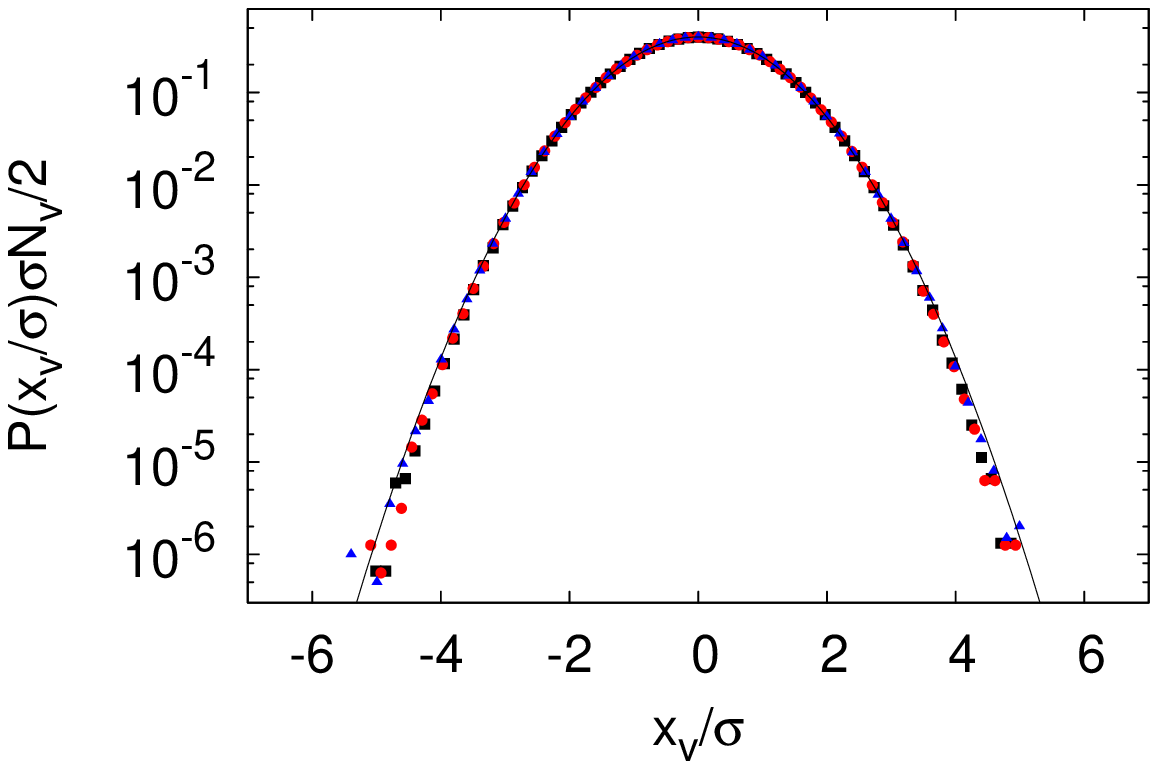}
 \includegraphics[width=.45\linewidth]{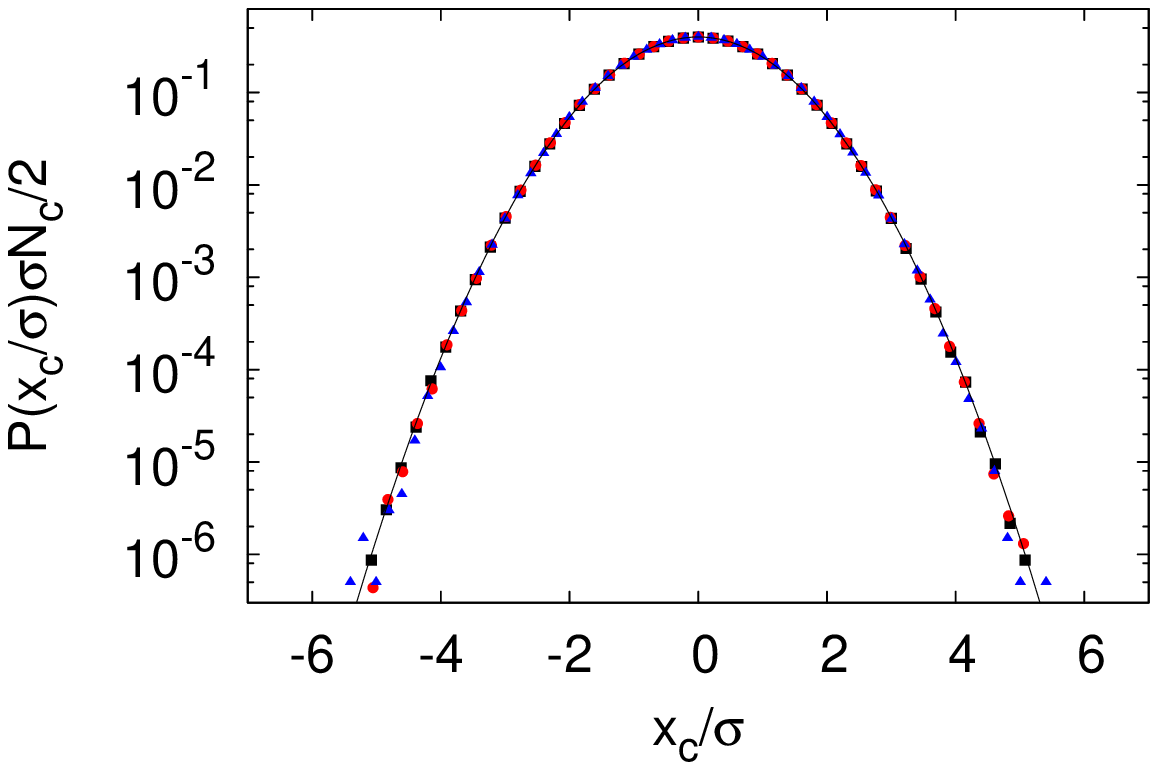}
 \caption{Scaled probability functions for a system of $100$ voters (left) and $100$ contrarians (right) for $h_c=h_v$ and $a_v/h_v=a_c/h_v=10^{-3}$ (black squares), $10^{-1}$ (red circles), and $10^{2}$ (blue triangle). For each set of data, it is $\sigma=\sqrt{\mean{x^2_{v,c}}}$. The solid line is the guassian distribution with zero mean and unit variance.}
 \label{fig:6}
\end{figure}

\section{Conclusions\label{sec:4}}
An agent-based model of voters and contrarians have been proposed and studied at the mean-field level (all-to-all interactions). The voters follow the noisy voter model dynamics, while the contrarians can change state by copying opposite states or by means of and intrinsic noise. The model is quite general, analytically tractable, and reduces to simpler ones: the (noisy) voter model (with and without zealots), the only-contrarians model (with and without zealots), and the two-role model, among others. 

Contrarians and voters behave differently if analyzed separately. The former always reach a steady state with a concave probability function (the bimodal phase), which in turns is very close to a Gaussian distribution. Moreover, the second moment of the distribution of contrarians is an increasing function of the noisy to anti-herding ratio, meaning that it is always smaller than that of system of noisy voters, the two quantities being coincident only when the intrinsic noises are the only source of transitions between opinions. 

In general, the presence of a small amount of contrarians modifies the dynamics of the voters critically. Particularly, if the number of contrarians is smaller than $4$, the noisy voters can be in the bimodal, plain, and unimodal phases, depending on the value of their intrinsic noise. For the number of contrarians bigger or equal to four, however, the phase transition disappears, and only the unimodal phase, both for the voters and contrarians, prevails. The situation is similar to that described by Galam using his majority model \cite{ga04}, the main difference happening when the number of contrarians is below the critical value: the bimodal phase of our model preserves the up-down (optimistic-pessimistic) symmetry, while the Galam model reaches a state with the fractions of agents with different opinions being different. 

When the numbers of voters and contrarians are of the same order, the probability distribution of contrarians is almost a Gaussian distribution. In addition, if the voter noise and herding coefficients are of the same order, the dynamics of voters and contrarians decouple, although their respective transition rates for voters and contrarians depend on the number of agents of each kind. In this latter case, the probability function of voters is also very close to the Gaussian distribution. 

\appendix

\section{Second moments \label{ap1}}

The exact expressions for the steady-state second moments of a general mixture of voters and contrarians are:
\begin{eqnarray}
  \label{eq:ap1}
  \nonumber 
  \mean{x_v^2}&=&[(2 a_v + h_v) N_v (-h_c (2 a_v + 2 a_c + h_v + h_c) + (2 a_c + h_c) (2 (a_v + a_c) + h_c) N_v)\\ \nonumber &&+ (-4 a_v^2 h_c - 2 (2 a_v + h_v) h_c (h_v + h_c) - 2 a_c (2 a_v h_c + h_v (h_v + h_c))\\ \nonumber && + (4 a_c (4 a_v (a_v + a_c) + (3 a_v + 2 a_c) h_v + h_v^2) + (12 a_v (a_v + 2 a_c) + 2 (5 a_v + 6 a_c) h_v + 3 h_v^2) h_c \\ \nonumber && \qquad + 4 (2 a_v + h_v) h_c^2) N_v) N_c + 2 (2 a_v + h_v) (a_c + h_c) (2 a_v + 2 a_c + h_v + 2 h_c) N_c^2]\\ \nonumber && \times [N_v (4 a_v^2 (N_v + N_c) (2 a_c (N_v + N_c) + h_c (-1 + N_v + 2 N_c)) \\ \nonumber && + 2 a_v ((2 a_c (N_v + N_c) + h_c (-1 + N_v + 2 N_c)) (2 a_c (N_v + N_c) + h_c (N_v + 2 N_c)) \\ \nonumber && + h_v (2 a_c (N_v + N_c) (1 + 2 N_c) + h_c (-1 + 3 N_v N_c + 4 N_c^2))) \\ \nonumber &&  + h_v (4 a_c^2 (1 + N_c) (N_v + N_c) + h_c (h_c (1 + 2 N_c) (-1 + N_v + 2 N_c) + h_v (-1 + N_c + 2 N_c^2)) \\ && \qquad + 2 a_c (h_v N_c (1 + N_c) + h_c (-1 + 2 N_v + 2 N_c + 3 N_v N_c + 4 N_c^2))))]^{-1},
\end{eqnarray}
\begin{eqnarray}
  \label{eq:ap2}
  \nonumber
  \mean{x_c^2}&=&[4 a_v^2 (2 a_c + h_c) (N_v + N_c)^2 + h_v (4 a_c^2 (1 + N_c) (N_v + N_c) + h_c N_c (h_v + h_c + 3 h_c N_v + h_v N_c + 2 h_c N_c) \\ \nonumber &&+ 2 a_c (h_v N_c (1 + N_c) + h_c (N_v + N_c) (2 + 3 N_c))) \\ \nonumber && + 2 a_v (4 a_c^2 (N_v + N_c)^2 + 2 a_c (N_v + N_c) (h_v + 2 h_c N_v + 2 h_v N_c + 3 h_c N_c) \\ \nonumber && \qquad + h_c (h_v (N_v + N_c) (1 + 2 N_c) + h_c ((-1 + N_v) N_v + 4 N_v N_c + 2 N_c^2)))] \\ \nonumber && \times [N_c (4 a_v^2 (N_v + N_c) (2 a_c (N_v + N_c) + h_c (-1 + N_v + 2 N_c)) \\ \nonumber && + 2 a_v ((2 a_c (N_v + N_c) + h_c (-1 + N_v + 2 N_c)) (2 a_c (N_v + N_c) + h_c (N_v + 2 N_c)) \\ \nonumber && + h_v (2 a_c (N_v + N_c) (1 + 2 N_c) + h_c (-1 + 3 N_v N_c + 4 N_c^2))) \\ \nonumber && + h_v (4 a_c^2 (1 + N_c) (N_v + N_c) + h_c (h_c (1 + 2 N_c) (-1 + N_v + 2 N_c) + h_v (-1 + N_c + 2 N_c^2)) \\ && + 2 a_c (h_v N_c (1 + N_c) + h_c (-1 + 2 N_v + 2 N_c + 3 N_v N_c + 4 N_c^2))))]^{-1},
\end{eqnarray}
\begin{eqnarray}
  \label{eq:ap3}
\nonumber
  \mean{x_vx_c}&=&[2 a_c (h_v^2 (1 + N_c) + h_v (2 a_v - h_c) (N_v + N_c) - 2 a_v h_c (N_v + N_c)) \\ \nonumber &&+ h_c (h_v^2 (1 + N_c) - 2 a_v h_c (-1 + N_v + 2 N_c) + h_v (2 a_v (N_v + N_c) - h_c (-1 + N_v + 2 N_c)))]\\ \nonumber && \times [4 a_v^2 (N_v +N_c) (2 a_c (N_v + N_c) + h_c (-1 + N_v + 2 N_c)) \\ \nonumber &&+ 2 a_v ((2 a_c (N_v + N_c) + h_c (-1 + N_v + 2 N_c)) (2 a_c (N_v + N_c) + h_c (N_v + 2 N_c)) \\ \nonumber &&+ h_v (2 a_c (N_v + N_c) (1 + 2 N_c) + h_c (-1 + 3 N_v N_c + 4 N_c^2))) + h_v (4 a_c^2 (1 + N_c) (N_v + N_c) \\ \nonumber &&+ h_c (h_c (1 + 2 N_c) (-1 + N_v + 2 N_c) + h_v (-1 + N_c + 2 N_c^2)) \\ &&+ 2 a_c (h_v N_c (1 + N_c) + h_c (-1 + 2 N_v + 2 N_c + 3 N_v N_c + 4 N_c^2)))]^{-1}.
\end{eqnarray}
 
\section*{Acknowledgements}
We acknowledge financial support from Agencia Estatal de Investigaci\'on (AEI, Spain) and Fondo Europeo de Desarrollo Regional under Project ESoTECoS Grant No. FIS2015-63628-C2-2-R (AEI/FEDER,UE). 
\section*{References}

\bibliography{mybibfile}

\end{document}